\documentclass[12pt,a4paper]{article}
\usepackage{amsmath,amsthm,amsfonts,amssymb,bbm,cite}
\usepackage{psfrag,graphicx}

\newtheorem{prop}{Proposition}
\newtheorem{thm}[prop]{Theorem}

\newtheorem{lem}[prop]{Lemma}
\newtheorem{defin}[prop]{Definition}

\newtheorem{rem}[prop]{Remark}
\newenvironment{remark}{\begin{rem}\normalfont}{\end{rem}}

\newcommand{\Id}{\mathbbm{1}}
\newcommand{\Or}{\mathcal{O}}
\newcommand{\Z}{\mathbb{Z}}
\newcommand{\E}{\mathbb{E}}
\newcommand{\R}{\mathbb{R}}
\newcommand{\Pb}{\mathbb{P}}
\newcommand{\I}{{\rm i}}
\newcommand{\dx}{{\rm d}}

\renewcommand{\Re}{{\rm Re}}
\renewcommand{\Im}{{\rm Im}}

\DeclareMathOperator*{\Tr}{Tr}

\title{\textbf{Dimers and orthogonal polynomials:\\ connections with random matrices}\\[0.5em]{\large Extended lecture notes of the minicourse at IHP}}
\author{\textbf{Patrik L. Ferrari}\\ {Bonn University} \\ {\small \texttt{ferrari@uni-bonn.de}}}
\date{}

\begin{document}
\maketitle
\sloppy

\vfill
\begin{abstract}
In these lecture notes we present some connections between random matrices, the asymmetric exclusion process, random tilings. These three apparently unrelated objects have (sometimes) a similar mathematical structure, an interlacing structure, and the correlation functions are given in terms of a kernel. In the basic examples, the kernel is expressed in terms of orthogonal polynomials.
\end{abstract}
\vfill

\newpage

\tableofcontents

\newpage

\section{Structure of these lecture notes}
In these notes we explain why there are limit processes and distribution functions which arise in random matrix theory, interacting particle systems, stochastic growth models, and random tilings models. This is due to a common mathematical structure describing special models in the different fields. In Section~\ref{SectGUE} we introduce the mathematical structure in the context of the Gaussian Unitary Ensemble and its eigenvalues' minor process. In Section~\ref{sectTASEP} we introduce the totally asymmetric simple exclusion process (TASEP), a particles' process sharing the same structure as the GUE minor process. Finally, in Section~\ref{sect2plus1growth} we discuss the extension of TASEP to an interacting particle system in $2+1$ dimensions. This model has two projections which are still Markov processes, which are (1) TASEP and (2) a discrete space analogue of Dyson's Brownian motion of random matrices. Also, projections at fixed time of the model leads to random tilings measures, one of which is the measure arising from the well known shuffling algorithm for the Aztec diamond.

Some books in random matrix theory are~\cite{Meh91,AGZ10,For10}. In the handbook~\cite{RMT11} one finds a lot of applications of random matrices and related models. For instance, the relation of random matrices with growth models is discussed in~\cite{FS10}, while determinantal point processes in~\cite{BorReview11}.

\section{Gaussian Unitary Ensemble of random matrices (GUE)}\label{SectGUE}

\subsection{The Gaussian Ensembles of random matrices}\label{subsectGUE}
The Gaussian ensembles of random matrices have been introduced by physicists (Dyson, Wigner, ...) in the sixties to model statistical properties of the resonance spectrum of heavy nuclei. The energy levels of a quantum system are the eigenvalues of a Hamiltonian. For heavy nuclei some properties of the spectrum, like eigenvalues' spacing statistics, seemed to have some regularity (e.g., repulsions) common for all heavy nuclei. In other words there was some universal behavior. If such a universal behavior exists, then it has to be the same as the behavior of choosing a \emph{random Hamiltonian}. Moreover, since the heavy nuclei have a lot of bound states (i.e., where the eigenvalues with normalizable eigenfunctions), the idea is to approximate the Hamiltonian by a \emph{large matrix with random entries}.

Even assuming the entries of the matrix random, to have a chance to describe the physical properties of heavy atoms, the matrices need to satisfy the intrinsic symmetries of the systems:
\begin{enumerate}
  \item a \emph{real symmetric} matrix can describe a system with \emph{time reversal symmetry} and \emph{rotation invariance or integer magnetic momentum},
  \item a \emph{real quaternionic} matrix (i.e., the basis are the Pauli matrices) can be used for \emph{time reversal symmetry} and \emph{half-integer magnetic momentum},
  \item a \emph{complex hermitian} matrix can describe a system which is \emph{not time reversal invariant} (e.g., with external magnetic field).
\end{enumerate}

We now first focus on the complex hermitian matrices case, since it is the one we are going to discuss later on.
\begin{defin}\label{defGUE}
The \textbf{Gaussian Unitary Ensemble} (GUE) of random matrices is a probability measure $\Pb$ on the set of $N\times N$ complex hermitian matrices given by
\begin{equation}\label{eqGUE}
\Pb(\dx H)=\frac{1}{Z_N} \exp\left(-\frac{\beta}{4N} \Tr(H^2)\right)  \dx H,\quad \textrm{with }\beta=2,
\end{equation}
where $\dx H=\prod_{i=1}^N \dx H_{i,i} \prod_{1\leq i<j\leq N}\dx\Re(H_{i,j})\dx\Im(H_{i,j})$ is the reference measure, and $Z_N$ is the normalization constant.
\end{defin}

The meaning of $\beta=2$ will be clear once we consider the induced measure on the eigenvalues. The name GUE is due to the Gaussian form of the measure (\ref{eqGUE}) and its invariance over the unitary transformations. This invariance is physically motivated, for systems which do not depend on the choice of basis used to describe them. By imposing that the measure $\Pb$ is (a) invariant under the change of basis (in the present case, invariant under the action of the group of symmetry $U(N)$) and (b) the entries of the matrices as independent random variables (of course, up to the required symmetry), then the only solutions are measure of the form
\begin{equation}\label{eqGUEgen}
\exp\left(-a \Tr(H^2)+b \Tr(H)+c\right), \quad a>0,b,c\in\R.
\end{equation}
The value of $c$ is determined by the normalization requirement, while by an appropriate shift of the zero of the energy (i.e., $H\to H-E$ for some given $E$), we can set $b=0$. The energy shift is irrelevant from the physical point of view, because by the first principle of thermodynamics the energy of a system is an extensive observable defined \emph{up to a constant}. The value of $a$ remains free to be chosen. Different choices can be easily compared since they are just a change of scale of each others. In the literature there are mainly three typical choices, see Table~\ref{Table1}, each one with its own reason.
\begin{table}
\begin{tabular}{|l|c|c|c|}
  \hline
   & $a=1/2N$ & $a=1$ & $a=N$ \\
  \hline
  & & & \\[-0.9em]
  Largest eigenvalue & $2N+\Or(N^{1/3})$ & $\sqrt{2N}+\Or(N^{-1/6})$ & $\sqrt{2}+\Or(N^{-2/3})$ \\
  & & & \\[-0.9em]
  Eigenvalues density & $\Or(1)$ & $\Or(N^{1/2})$ & $\Or(N)$ \\
  \hline
\end{tabular}
\caption{Typical scaling for the Gaussian Unitary Ensemble}
\label{Table1}
\end{table}

Another way to obtain (\ref{eqGUE}) is to take the random variables, $H_{i,i}\sim {\cal N}(0,N)$ for $i=1,\ldots,N$, and $\Re(H_{i,j})\sim {\cal N}(0,N/2)$, $\Im(H_{i,j})\sim {\cal N}(0,N/2)$ for $1\leq i<j\leq N$ as independent random variables.

For the class of real symmetric (resp.\ quaternionic), one defines the Gaussian Orthogonal Ensemble (GOE) (resp.\ Gaussian Symplectic Ensemble (GSE)) as in Definition~\ref{defGUE} but with $\beta=1$ (resp.\ $\beta=4$) and, of course, the reference measure is now the product Lebesgue measure over the independent entries of the matrices.

\subsection{Eigenvalues' distribution}\label{subsectEigenDistr}
One quantity of interest for random matrices are the eigenvalues, their distributions and correlations. For the Gaussian ensembles of random matrices the eigenvalues are independent of the choice of basis. This allows to explicitly compute the projection of the random matrix measure to the eigenvalues measure with the following result. Denote by $P_{\rm GUE}(\lambda)$ the probability density of eigenvalues at $\lambda \in \R^N$.

\begin{prop}
Let $\lambda_1,\lambda_2,\ldots,\lambda_N\in \R$ denote the $N$ eigenvalues of a random matrix $H$ with measure (\ref{eqGUE}). Then, the joint distribution of eigenvalues is given by
\begin{equation}\label{eqGUEeigenvals}
P_{\rm GUE}(\lambda)=\frac{1}{Z_N} |\Delta_N(\lambda)|^\beta \prod_{i=1}^N \exp\left(-\frac{\beta}{4N}\lambda_i^2\right), \quad\textrm{with }\beta=2,
\end{equation}
$\Delta_N(\lambda):=\prod_{1\leq i<j\leq N}(\lambda_j-\lambda_i)$ is the Vandermonde determinant, and $Z_N$ is a normalization constant.
\end{prop}
$\Delta_N$ is called a determinant because it holds
\begin{equation}
\Delta_N(\lambda) =\det\left[\lambda_i^{j-1}\right]_{1\leq i,j\leq N}.
\end{equation}
Notice that $P_{\rm GUE}(\#\textrm{ e.v.}\in[x,x+\dx x])\sim (\dx x)^2$, so that the probability of having double eigenvalues is zero: it is a \textbf{simple point process}.

For GOE (resp.\ GSE) the joint distributions of eigenvalues has the form~(\ref{eqGUEeigenvals}) but with $\beta=1$ (resp.\ $\beta=4$) instead.

\subsection{Orthogonal polynomials}\label{subsectOrthoPoly}
The correlation function for GUE eigenvalues can be described using Hermite orthogonal polynomials. Therefore, let us shortly discuss orthogonal polynomials on $\R$. Formulas can be easily adapted for polynomials on $\Z$ by replacing the Lebesgue measure by the counting measure and integrals by sums.
\begin{defin}\label{defOrthoPoly}
Given a weight $\omega:\R\mapsto \R_+$, the \textbf{orthogonal polynomials} $\{q_k(x),k\geq 0\}$ are defined by the following two conditions:
\begin{enumerate}
  \item $q_k(x)$ is a polynomial of degree $k$ with $q_k(x)=u_k x^k+\ldots$, $u_k>0$,
  \item the $q_k(x)$ satisfy the orthonormality condition,
  \begin{equation}\label{eqOrthonormality}
    \langle q_k,q_l \rangle_\omega:=\int_{\R} \omega(x) q_k(x) q_l(x) \dx x=\delta_{k,l}.
  \end{equation}
\end{enumerate}
\end{defin}

\begin{remark}
There are other normalizations which are often used, like in the \emph{Askey Scheme of hypergeometric orthogonal polynomials}~\cite{KS96}. Sometimes, the polynomials are taken to be \emph{monic}, i.e., $u_k=1$ and the orthonormality condition has then to be replaced by an orthogonality condition $\int_{\R} \omega(x) \tilde q_k(x) \tilde q_l(x) \dx x=c_k \delta_{k,l}$. Of course $\tilde q_k(x)=q_k(x)/u_k$ and $c_k=1/u_k^2$. Sometimes, the polynomials are neither orthonormal (like in Definition~\ref{defOrthoPoly}) nor monic, like the standard Hermite polynomials that we will encounter, and are given by derivatives of a generating function.
\end{remark}

A useful formula for sums of orthogonal polynomials is the \textbf{Christoffel-Darboux formula}:
\begin{equation}\label{eqChristDarb}
\sum_{k=0}^{N-1} q_k(x) q_k(y) = \left\{
\begin{array}{ll}
\displaystyle{\frac{u_{N-1}}{u_N}\frac{q_N(x) q_{N-1}(y)-q_{N-1}(x) q_N(y)}{x-y}},&\textrm{ for }x\neq y,\\[1em]
\displaystyle{\frac{u_{N-1}}{u_N} (q_N'(x) q_{N-1}(x)-q_{N-1}'(x) q_N(x))},&\textrm{ for }x=y.
\end{array}
\right.
\end{equation}
This formula is proven by employing the following three term relation
\begin{equation}\label{eq3TermRel}
q_n(x)=(A_n x+B_n)q_{n-1}(x)-C_n q_{n-2}(x),
\end{equation}
with $A_n>0, B_n, C_n>0$ some constants. See Appendix~\ref{appChristoffel} for details of the derivation. For polynomials as in Definition~\ref{defOrthoPoly}, it holds $A_n=u_n/u_{n-1}$ and $C_n=A_n/A_{n-1}=u_n u_{n-2}/u_{n-1}^2$.

\subsection{Correlation functions of GUE}\label{subsectGUEcorr}

Now we restrict to the GUE ensemble and discuss the derivation of the correlation functions for the GUE eigenvalues' point process.

Let the reference measure be Lebesgue. Then, the $n$-point correlation function, $\rho^{(n)}_{\rm GUE}(x_1,\ldots,x_n)$ is the probability density of finding \emph{an} eigenvalue \emph{at each} of the $x_k$, $k=1,\ldots,n$. $P_{\rm GUE}$ defined in (\ref{eqGUEeigenvals}) is symmetric with respect the permutation of the variables, which directly implies the following result.
\begin{lem}\label{lemGUEcorrfct}
The $n$-point correlation function for GUE eigenvalues is given by
\begin{equation}\label{eqGUEcorrfct}
\rho^{(n)}_{\rm GUE}(x_1,\ldots,x_n)=\frac{N!}{(N-n)!}\int_{\R^{N-n}} P_{\rm GUE}(x_1,\ldots,x_N) \dx x_{n+1} \ldots \dx x_N
\end{equation}
for $n=1,\ldots,N$ and $\rho^{(n)}_{\rm GUE}(x_1,\ldots,x_n)=0$ for $n>N$.
\end{lem}
It is important to notice that it is not fixed which eigenvalue is at which position. In particular, we have
\begin{equation}
\rho^{(1)}_{\rm GUE}(x)=\textrm{eigenvalues' density at }x
\end{equation}
that implies $\int_\R \rho^{(1)}_{\rm GUE}(x)\dx x=N$ and not $1$ (so, $\rho^{(1)}_{\rm GUE}(x)$ is not the density of a distribution function). More generally,
\begin{equation}
\int_{\R^n}\rho^{(n)}_{\rm GUE}(x_1,\ldots,x_n)\dx x_1\ldots \dx x_n=\frac{N!}{(N-n)!}.
\end{equation}

Our next goal is to do the integration in (\ref{eqGUEcorrfct}). For any family of polynomials \mbox{$\{q_k,k=0,\ldots,N-1\}$}, where $q_k$ has degree $k$, by multi-linearity of the determinant, we have
\begin{equation}
\Delta_N(\lambda)=\det[\lambda_i^{j-1}]_{1\leq i,j\leq N} = {\rm const}\times \det[q_{j-1}(\lambda_i)]_{1\leq i,j\leq N}.
\end{equation}
Therefore, setting $\omega(x):=\exp(-x^2/2N)$, we have
\begin{equation}\label{eq13}
\begin{aligned}
& P_{\rm GUE}(\lambda_1,\ldots,\lambda_N) \\ &= {\rm const}\times \det[q_{k-1}(\lambda_i)]_{1\leq i,k\leq N}  \det[q_{k-1}(\lambda_j)]_{1\leq k,j\leq N} \prod_{i=1}^N \omega(\lambda_i) \\
&={\rm const}\times \det\left[\sum_{k=1}^N q_{k-1}(\lambda_i) q_{k-1}(\lambda_j)\right]_{1\leq i,j\leq N} \prod_{i=1}^N \omega(\lambda_i).
\end{aligned}
\end{equation}
Notice that until this point, the polynomials $q$'s do not have to be orthogonal. However, if we choose the polynomials orthogonal with respect to the weight $\omega$, then the integrations in (\ref{eqGUEcorrfct}) becomes particularly simple and nice.
\begin{prop}\label{propGUEcorrfcf}
Let $q_k$ be orthogonal polynomials with respect to the weight $\omega(x)=\exp(-x^2/2N)$. Then,
\begin{equation}\label{eqGUEcorrfctDetForm}
\rho^{(n)}_{\rm GUE}(x_1,\ldots,x_n)=\det\left[K_N^{\rm GUE}(x_i,x_j)\right]_{1\leq i,j\leq n},
\end{equation}
where
\begin{equation}\label{eq14}
K_N^{\rm GUE}(x,y)=\sqrt{\omega(x)}\sqrt{\omega(y)} \sum_{k=0}^{N-1} q_k(x) q_k(y).
\end{equation}
\end{prop}

The proof of Proposition~\ref{propGUEcorrfcf} can be found in Appendix~\ref{AppGUEcorr}. What we need to do is to integrate out the $N-n$ variables one after the other and see that the determinant keeps the same entries but becomes smaller. For that we use the following two identities
\begin{equation}\label{eq16}
\begin{aligned}
\int_{\R}K_N^{\rm GUE}(x,x)\dx x&=N,\\
\int_{\R}K_N^{\rm GUE}(x,z)K_N^{\rm GUE}(z,y)\dx z &= K_N^{\rm GUE}(x,y),
\end{aligned}
\end{equation}
which holds precisely because the $q_k$'s in (\ref{eq14}) are the orthogonal polynomials with respect to $\omega(x)$. Correlation functions of the form (\ref{eqGUEcorrfctDetForm}) are called determinantal.
\begin{defin}\label{defDetPP}
A point process (i.e., a random point measure) is called \textbf{determinantal} if its $n$-point correlation function has the form
\begin{equation}
\rho^{(n)}(x_1,\ldots,x_n)=\det[K(x_i,x_j)]_{1\leq i,j\leq n}
\end{equation}
for some (measurable) function $K:\R^2\to \R$, called the \textbf{kernel} of the determinantal point process.
\end{defin}

One might ask when a measure defines a determinantal point process. A sufficient condition is the following (see Proposition~2.2 of~\cite{Bor98}, see also~\cite{TW98} for the GUE case).
\begin{thm}\label{thmBoro}
Consider a probability measure on $\R^N$ of the form
\begin{equation}\label{eqPropDPP}
\frac{1}{Z_N}\det[\Phi_i(x_j)]_{1\leq i,j\leq N} \det[\Psi_i(x_j)]_{1\leq i,j\leq N}\prod_{i=1}^N \omega(x_i) \dx x_i,
\end{equation}
with the normalization $Z_N\neq 0$. Then (\ref{eqPropDPP}) defines a determinantal point process with kernel
\begin{equation}
K_N(x,y)=\sum_{i,j=1}^N\Psi_i(x)[A^{-1}]_{i,j}\Phi_j(y),
\end{equation}
where $A=[A_{i,j}]_{1\leq i,j\leq N}$,
\begin{equation}
A_{i,j}=\langle \Phi_i,\Psi_j\rangle_\omega = \int_{\R} \omega(z) \Phi_i(z) \Psi_j(z)\dx z.
\end{equation}
\end{thm}

\subsection{GUE kernel and Hermite polynomials}\label{subsectGUEkernel}
The GUE kernel $K_N^{\rm GUE}$ can be expressed in terms of the \textbf{standard Hermite polynomials}, $\{H_n, n\geq 0\}$, defined by
\begin{equation}
H_k(y)=(-1)^k e^{y^2}\frac{\dx^k}{\dx y^k} e^{-y^2}.
\end{equation}
They satisfy
\begin{equation}
\int_{\R} e^{-y^2} H_k(y) H_l(y)\dx y= \sqrt{\pi} 2^k k! \delta_{k,l},
\end{equation}
with $H_k(y)=2^k y^k+\ldots$, and also
\begin{equation}
\frac{\dx}{\dx y}\left(e^{-y^2} H_n(y)\right)=-e^{-y^2}H_{n+1}(y)
\end{equation}
implies
\begin{equation}\label{eqRecRelHermitePol}
\int_{-\infty}^x e^{-y^2}H_{n+1}(y)\dx y=-e^{-x^2} H_n(x).
\end{equation}

By the change of variable $y=x/\sqrt{2N}$ and a simple computation, one shows that
\begin{equation}\label{eq23a}
q_k(x)=\frac{1}{\sqrt[4]{2\pi N}}\frac{1}{\sqrt{2^k k!}} H_k\left(\frac{x}{\sqrt{2N}}\right)
\end{equation}
are orthogonal polynomials with respect to $\omega(x)=\exp(-x^2/2N)$, and that
$u_k=(2\pi N)^{-1/4} k!^{-1/2} N^{-k/2}$. Then, Christoffel-Darboux formula~(\ref{eqChristDarb}) gives
\begin{equation}
K_N^{\rm GUE}(x,y)=
 \left\{
\begin{array}{ll}
\displaystyle{\frac{q_N(x)q_{N-1}(y)-q_{N-1}(x)q_N(y)}{x-y}} N e^{-(x^2+y^2)/4N},&\textrm{ for }x\neq y,\\[1em]
\displaystyle{(q_N'(x) q_{N-1}(x)-q_{N-1}'(x) q_N(x))} N e^{-(x^2+y^2)/4N},&\textrm{ for }x=y.
\end{array}
\right.
\end{equation}

With our normalization in (\ref{eqGUE}) the eigenvalues' density remains bounded and the largest eigenvalue is close to the value $2N$. Indeed, the eigenvalues' density at position $\mu N$ is given by
\begin{equation}\label{eqWignerSemicircle}
\rho^{(1)}(\mu N)=K_N^{\rm GUE}(\mu N,\mu N)\stackrel{N\to\infty}{\longrightarrow}
\left\{
\begin{array}{ll}
\displaystyle{\frac{1}{\pi}\sqrt{1-(\mu/2)^2}},&\textrm{ for }\mu\in[-2,2],\\[1em]
0,&\textrm{ otherwise}.
\end{array}
\right.
\end{equation}
The large-$N$ asymptotic density in (\ref{eqWignerSemicircle}) is called \textbf{Wigner's semicircle law}.

\subsection{Distribution of the largest eigenvalue: gap probability}\label{subsectGapProba}
The next question is how to compute the distribution of the largest eigenvalue, $\lambda_{\rm max}$. One uses the following simple relation,
\begin{equation}
\Pb(\lambda_{\rm max}\leq s)=\Pb(\textrm{no eigenvalue lies in }(s,\infty)),
\end{equation}
which is a special case of \textbf{gap probability}, i.e., probability that in a Borel set $B$ there are no eigenvalues. The gap probabilities are expressed in terms of $n$-point correlation functions as follows:
\begin{equation}\label{eqFredDet}
\begin{aligned}
&\Pb(\textrm{no eigenvalue lies in }B)=\E\bigg(\prod_{i}(1-\Id_B(\lambda_i))\bigg) \\
&=\sum_{n\geq 0}(-1)^n\E\bigg(\sum_{i_1<\ldots<i_n}\prod_{k=1}^n\Id_B(\lambda_{i_k})\bigg)
\stackrel{\rm sym}{=}\sum_{n\geq 0}\frac{(-1)^n}{n!}\E\bigg(
\sum_{\begin{subarray}{c}i_1,\ldots,i_n\\ \textrm{all different}\end{subarray}}\prod_{k=1}^n\Id_B(\lambda_{i_k})\bigg)\\
&=\sum_{n\geq 0}\frac{(-1)^n}{n!}\int_{B^n}\rho^{(n)}(x_1,\ldots,x_n) \, \dx x_1\ldots \dx x_n,
\end{aligned}
\end{equation}
where $\Id_B(x)=1$ if $x\in B$ and $\Id_B(x)=0$ if $x\not\in B$. The last step holds for simple point processes, which are point processes for which the probability of double occurrence of points (here eigenvalues) is zero.

For a determinantal point process, like the GUE eigenvalues we have
\begin{equation}\label{eq28}
\begin{aligned}
\Pb(\lambda_{\rm max}\leq s)&=\sum_{n=0}^{\infty}\frac{(-1)^n}{n!}\int_{(s,\infty)^n}\det[K_N^{\rm GUE}(x_i,x_j)]_{1\leq i,j\leq n}\dx x_1\ldots \dx x_n \\
&\equiv \det(\Id-K_N^{\rm GUE})_{L^2((s,\infty),\dx x)}.
\end{aligned}
\end{equation}
The series expansion in (\ref{eq28}) is called the Fredholm series expansion of the Fredholm determinant\footnote{If $M$ is a $n\times n$ matrix with eigenvalues $\mu_1,\ldots,\mu_n$, then $\det(\Id-M)=\prod_{j=1}^n (1-\mu_j)$. A Fredholm determinant is a generalisation of it for integral operators $\cal K$ with kernel $K$. See e.g.~\cite{Sim00,RS78III} for details.} $\det(\Id-K_N^{\rm GUE})_{L^2((s,\infty),\dx x)}$. In our case the sum over $n$ is actually finite because the kernel has finite rank. Indeed, for $n>N$ the correlation functions are equal to zero, since the kernel $K_N^{\rm GUE}$ has rank $N$. Here we kept the formulation of the general case.

\subsection{Correlation functions of GUE minors: interlacing structure}\label{subsectGUEminors}
In this section we explain how the determinantal structure holds in the enlarged setting of eigenvalues of minors. This extension is different from then one Eynard-Mehta formula~\cite{EM97,FN98} for Dyson's Brownian motion, see also~\cite{Jo03b} for a generic statement (the analogue of Theorem~\ref{ThmSasa} below).

Consider a $N\times N$ GUE random matrix $H$ and denote by $\lambda_1^N,\ldots,\lambda_N^N$ its eigenvalues. Denote by $H_m$ the $m\times m$ minor of the matrix $H$ where only the first $m$ rows and columns are kept. Let $\lambda_1^m,\ldots,\lambda_m^m$ be the eigenvalues of $H_m$. In~\cite{JN06,FN08b} the correlation functions of $\{\lambda_k^m,1\leq k\leq m\leq N\}$ are determined. It turns out that also in that case the correlation functions are determinantal (on $\{1,\ldots,N\}\times\R$).
\begin{figure}
\begin{center}
\psfrag{l11}[cc]{$\lambda_1^1$}
\psfrag{l12}[cc]{$\lambda_1^2$}
\psfrag{l22}[cc]{$\lambda_2^2$}
\psfrag{l13}[cc]{$\lambda_1^3$}
\psfrag{l23}[cc]{$\lambda_2^3$}
\psfrag{l33}[cc]{$\lambda_3^3$}
\psfrag{l14}[cc]{$\lambda_1^4$}
\psfrag{l24}[cc]{$\lambda_2^4$}
\psfrag{l34}[cc]{$\lambda_3^4$}
\psfrag{l44}[cc]{$\lambda_4^4$}
\psfrag{n}[cc]{$n$}
\psfrag{R}[cc]{$\R$}
\psfrag{1}[cc]{$1$}
\psfrag{2}[cc]{$2$}
\psfrag{3}[cc]{$3$}
\psfrag{4}[cc]{$4$}
\includegraphics[height=4cm]{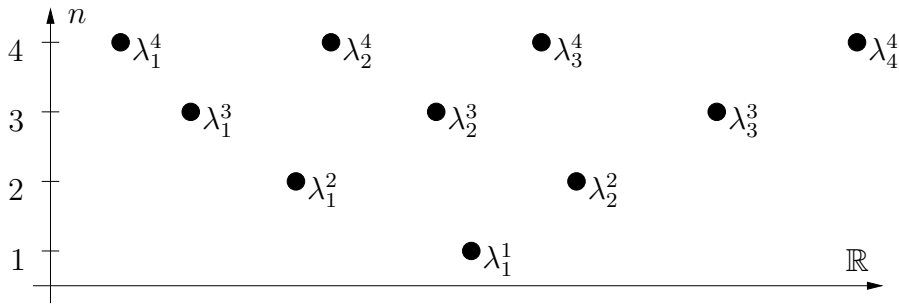}
\caption{Interlacing structure of the GUE minors' eigenvalues.}
\label{FigGUEminors}
\end{center}
\end{figure}

Let us order the eigenvalues for each minor as \mbox{$\lambda_1^m\leq \lambda_2^m\leq \ldots\leq \lambda_m^m$}. Then, the GUE minor measure can be written as, see e.g.~\cite{FN08b},
\begin{equation}\label{eqMeasureGUEminorsB}
{\rm const}\times \bigg(\prod_{m=1}^{N-1} \Id(\lambda^{m}\prec\lambda^{m+1})\bigg) \det[\Psi^N_{N-i}(\lambda_j^N)]_{1\leq i,j\leq N},
\end{equation}
where \begin{equation}
\Psi^N_{N-k}(x)=\frac{(-1)^{N-k}}{(2N)^{(N-k)/2}}e^{-x^2/2N} H_{N-k}(x/\sqrt{2N}),
\end{equation}
and $\lambda^{m}\prec\lambda^{m+1}$ means that the eigenvalues' configuration satisfies the \textbf{interlacing condition}
\begin{equation}\label{eqInterlacingGUE}
\lambda_1^{m+1}<\lambda_1^{m}\leq \lambda_2^{m+1} < \lambda_2^{m}\leq \ldots<\lambda_{m}^{m} \leq \lambda_{m+1}^{m+1},
\end{equation}
see Figure~\ref{FigGUEminors} for an illustration. Strictly speaking, one should not have strict inequality, but this is irrelevant since the events with $\lambda_k^n=\lambda_k^{n+1}$ have probability zero.

It is easy to verify that
\begin{equation}\label{eq32b}
\int_{\R^{n(n-1)/2}} \prod_{1\leq k \leq m\leq n-1} \dx\lambda_k^m \prod_{m=1}^{n-1} \Id(\lambda^{m}\prec\lambda^{m+1}) = \Delta_n(\lambda_1^n,\ldots,\lambda_n^n).
\end{equation}
This means that summing over the $\lambda_k^n$, $1\leq k \leq n \leq N-1$ we recover a measure as in (\ref{eqPropDPP}), with $\Psi_k$ replaced by $\Psi^N_k$ and $\Phi_k$ a polynomial of degree~$k$.

In the same spirit as in Eynard-Mehta formula, it turns out to be convenient to write the indicator function over interlacing configurations as a determinant. Here, however, the sets $\{\lambda_j^m,1\leq j \leq m\}$ and $\{\lambda_j^{m+1},1\leq j \leq m+1\}$ have different sizes. To keep notations compact, we will introduce the symbol $\lambda_{m+1}^m={\rm virt}$. We call them virtual variables, since they are not eigenvalues of a matrix.
Defining $\phi(x,y)=\Id(x>y)$, $\phi({\rm virt},y)=1$, then
\begin{equation}
\det[\phi(\lambda_i^{m},\lambda_j^{m+1})]_{1\leq i,j\leq m+1}=\left\{
\begin{array}{ll}
1,&\textrm{if (\ref{eqInterlacingGUE}) is satisfied},\\
0,&\textrm{otherwise}.
\end{array}
\right.
\end{equation}
Therefore the measure on the GUE eigenvalues' minor is given by
\begin{equation}\label{eqMeasureGUEminors}
{\rm const}\times \bigg(\prod_{m=1}^{N-1} \det[\phi(\lambda_i^{m},\lambda_j^{m+1})]_{1\leq i,j\leq m+1}\bigg) \det[\Psi^N_{N-i}(\lambda_j^N)]_{1\leq i,j\leq N}.
\end{equation}
Until now the eigenvalues are still ordered for each minor. We can relax this constraint whenever we want (for instance to apply Theorem~\ref{ThmSasa} below). Indeed, the measure (\ref{eqMeasureGUEminors}) is symmetric under the permutation of the eigenvalues of a given minor. Thus relaxing the constraint it results only in a change of normalisation constant.

A measure of the form (\ref{eqMeasureGUEminors}) has determinantal correlations~\cite{BFPS06}. The difference with the above case is that the correlation functions are determinantal on $\{1,\ldots,N\}\times\R$ instead of $\R$. This means the following. The probability density of finding an eigenvalue of $H_{n_i}$ at position $x_i$, for $i=1,\ldots,n$, is given by
\begin{equation}
\rho^{(n)}((n_i,x_i),1\leq i\leq n)=\det\left[K_N^{\rm GUE}(n_i,x_i;n_j,x_j)\right]_{1\leq i,j\leq n}.
\end{equation}

To explain the formula for the extended kernel $K_N^{\rm GUE}$ we need to introduce some definitions.
Let us set
\begin{equation}
\Psi^{n}_{n-k}(x):=(\phi*\Psi^{n+1}_{n+1-k})(x).
\end{equation}
Then, from (\ref{eqRecRelHermitePol}) we have
\begin{equation}
\Psi^n_{n-k}(x)=\frac{(-1)^{n-k}}{(2N)^{(n-k)/2}}e^{-x^2/2N} H_{n-k}(x/\sqrt{2N})
\end{equation}
for $1\leq k \leq n$. Next we need to find $\{\Phi^n_{n-k}(x),k=1,\ldots,n\}$ orthogonal, with respect to the weight $\omega(x)=1$, to the functions $\{\Psi^n_{n-j}(x),j=1,\ldots,n\}$, and such that
\begin{equation}
{\rm span}\{\Phi_0^n(x),\ldots,\Phi_{n-1}^n(x)\}={\rm span}\{1,x,\ldots,x^{n-1}\}.
\end{equation}
We find
\begin{equation}\label{eq35}
\Phi^n_{n-j}(x)=\frac{(-1)^{n-j}}{\sqrt{2\pi}(n-j)!}\left(\frac{N}{2}\right)^{(n-j)/2} H_{n-j}(x/\sqrt{2N}).
\end{equation}
Finally, let us define by $\phi^{*(n_2-n_1)}$ the convolution of $\phi$ with itself $n_2-n_1$ times, namely, for $n_2>n_1$,
\begin{equation}
(\phi^{*(n_2-n_1)})(x_1,x_2)=\frac{(x_2-x_1)^{n_2-n_1-1}}{(n_2-n_1-1)!}\Id_{[x_2-x_1\geq 0]}.
\end{equation}
Then applying Theorem~\ref{ThmSasa} to this particular case we obtain the following result.
\begin{prop}\label{propGUEminors}
With the above notations, the correlation functions of the GUE minors are determinantal with kernel given by
\begin{equation}
K_N^{\rm GUE}(n_1,x_1;n_2,x_2)=-(\phi^{*(n_2-n_1)})(x_1,x_2)\Id_{[n_1<n_2]}+\sum_{k=1}^{n_2} \Psi^{n_1}_{n_1-k}(x_1)\Phi^{n_2}_{n_2-k}(x_2).
\end{equation}
\end{prop}

This result is just a particular case of a more general statement. Consider a measure on $\{x_i^n,1\leq i\leq n\leq N\}$ of the form
\begin{equation}\label{Sasweight}
\frac{1}{Z_N}\bigg(\prod_{n=1}^{N-1} \det[\phi_n(x_i^n,x_j^{n+1})]_{1\leq i,j\leq n+1}\bigg) \det[\Psi_{N-i}^{N}(x_{j}^N)]_{1\leq i,j \leq N},
\end{equation}
where $x_{n+1}^n$ are some virtual variables and $Z_N$ is a normalization constant. It turns out that if $Z_N\neq 0$, then the correlation functions are determinantal. Define
\begin{equation}\label{Sasdef phi12}
\phi^{(n_1,n_2)}(x,y)=\left\{\begin{array}{ll}(\phi_{n_1}* \cdots *
\phi_{n_2-1})(x,y),& n_1<n_2,\\ 0,& n_1\geq n_2,\end{array}\right.
\end{equation}
where $(a* b)(x,y)=\int_{\R} a(x,z) b(z,y)\dx z$, and, for $1\leq n<N$,
\begin{equation}\label{Sasdef_psi}
\Psi_{n-j}^{n}(x) := (\phi^{(n,N)} * \Psi_{N-j}^{N})(y), \quad j=1,2,\ldots,N.
\end{equation}
Set $\phi_0(x_1^0,x)=1$. Then the functions
\begin{equation}
\{ (\phi_0*\phi^{(1,n)})(x_1^0,x), \dots,(\phi_{n-2}*\phi^{(n-1,n)})(x_{n-1}^{n-2},x), \phi_{n-1}(x_{n}^{n-1},x)\}
\end{equation}
are linearly independent and generate the $n$-dimensional space $V_n$. Further define a set of functions $\{\Phi_j^{n}(x), j=0,\ldots,n-1\}$ spanning $V_n$ defined by the orthogonality relations
\begin{equation}\label{Sasortho}
\int_\R \Phi_i^n(x) \Psi_j^n(x) \dx x= \delta_{i,j}
\end{equation}
for $0\leq i,j\leq n-1$.

\begin{thm}\label{ThmSasa}
Assume that we have a measure on $\{x_i^n,1\leq i\leq n\leq N\}$ given by (\ref{Sasweight}). If $Z_N\neq 0$, then the measure has determinantal correlations. Further, under
\begin{equation}
\textrm{Assumption (A):}\quad\phi_n(x_{n+1}^n,x)=c_n \Phi_0^{(n+1)}(x),\quad c_n\neq 0, \forall n=1,\ldots,N-1,
\end{equation}
the kernel has the simple form
\begin{equation}\label{SasK}
K(n_1,x_1;n_2,x_2)= -\phi^{(n_1,n_2)}(x_1,x_2)+ \sum_{k=1}^{n_2} \Psi_{n_1-k}^{n_1}(x_1) \Phi_{n_2-k}^{n_2}(x_2).
\end{equation}
\end{thm}
\begin{remark}
Without Assumption (A), the correlations functions are still determinantal but the formula is modified as follows. Let $M$ be the $N\times N$ dimensional matrix defined by $[M]_{i,j}=(\phi_{i-1}*\phi^{(i,N)}*\Psi^N_{N-j})(x_i^{i-1})$. Then
\begin{eqnarray}
& &K(n_1,x_1;n_2,x_2)\\
&=& -\phi^{(n_1,n_2)}(x_1,x_2)+ \sum_{k=1}^{N} \Psi_{n_1-k}^{n_1}(x_1) \sum_{l=1}^{n_2} [M^{-1}]_{k,l} (\phi_{l-1}*\phi^{(l,n_2)})(x_l^{l-1},x_2).\nonumber
\end{eqnarray}
Theorem~\ref{ThmSasa} is proven by using the framework of~\cite{RB04}.
\end{remark}
In the case of the measure (\ref{eqMeasureGUEminors}), the $n$-dimensional space $V_n$ is spanned by $\{1,x,\ldots,x^{n-1}\}$. This is a consequence of (\ref{eq32b}). Thus the $\Phi^n_{k}$ are polynomials of degree $k$, compare with (\ref{eq35}).

In the next section we are going to consider an interacting particle system, that is the setting where Theorem~\ref{ThmSasa} was discovered.

\newpage

\section{Totally Asymmetric Simple Exclusion Process (TASEP)}\label{sectTASEP}

\subsection{Continuous time TASEP: interlacing structure}\label{subsectTASEPinterlacing}
The totally asymmetric simple exclusion process (TASEP) is one of the simplest interacting stochastic particle systems. It consists of particles on the lattice of integers, $\Z$, with \emph{at most one particle at each site} (exclusion principle). The dynamics in continuous time is as follows. Particles jump on the neighboring right site with rate $1$ provided that the site is empty. This means that jumps are independent of each other and take place after an exponential waiting time with mean $1$, which is counted from the time instant when the right neighbor site is empty.

Here we consider all particles with equal rate $1$. However, the framework which we explain below, can be generalized to particle-dependent rates and also particles can jump on both directions: by jumping on their right, particles can be blocked, while on the left if a site is occupied, then it is pushed by the jumping particle. This generalization, called PushASEP, together with a partial extension to space-time correlations is the content of our paper~\cite{BF07}. We warn the reader that the resulting model is not the same as the well-studied partially asymmetric simple exclusion process, where also the left jumps are blocked if their left site is occupied.

On the macroscopic level the particle density, $u(x,t)$, evolves deterministically according to the Burgers equation $\partial_t u + \partial_x(u(1-u))=0$~\cite{Rez91}. Therefore a natural question is to focus on fluctuation properties, which exhibit rather unexpected features. The asymptotic results can be found in the literature, see Appendix~\ref{AppAsymptRef}. Here we focus on a method which can be used to analyze the joint distributions of particles' positions. This method is based on a interlacing structure first discovered by Sasamoto in~\cite{Sas05}, later extended and generalized in a series of papers, starting with~\cite{BFPS06}. We explain the key steps following the notations of~\cite{BFPS06}, where the details of the proofs can be found.

Consider the TASEP with $N$ particles starting at time $t=0$ at positions $y_N <\ldots < y_2 < y_1$. The first step is to obtain the probability that
at time $t$ these particles are at positions $x_N < \ldots < x_2 <x_1$, which we denote by
\begin{equation}
G(x_1,\ldots,x_N;t)=\Pb((x_N,\ldots,x_1;t)|(y_N,\ldots,y_1;0)).
\end{equation}
This function has firstly been determined using Bethe-Ansatz method~\cite{Sch97}. A posteriori, the result can be checked directly by writing the evolution equation for $G$ (also known as master equation).
\begin{lem}\label{lem1}
The transition probability is given by
\begin{equation}\label{eqGreen}
G(x_1,\ldots,x_N;t)=\det(F_{i-j}(x_{N+1-i}-y_{N+1-j},t))_{1\leq i,j\leq N}
\end{equation}
with
\begin{equation}\label{eqFn}
F_{n}(x,t)=\frac{(-1)^n}{2\pi \I} \oint_{\Gamma_{0,1}} \frac{\dx w}{w}
\frac{(1-w)^{-n}}{w^{x-n}}e^{t(w-1)},
\end{equation}
where $\Gamma_{0,1}$ is any simple loop oriented anticlockwise which
includes $w=0$ and $w=1$.
\end{lem}
The key property of Sasamoto's decomposition is the following relation
\begin{equation}\label{eqRecRel}
F_{n+1}(x,t)=\sum_{y\geq x} F_n(y,t).
\end{equation}
Denote $x_1^k:=x_k$ the position of TASEP particles. Using the multi-linearity of the determinant and (\ref{eqRecRel}) one obtains
\begin{equation}\label{eqLem2}
G(x_1,\ldots,x_N;t)=\sum_{\cal D'}
\det(F_{-j+1}(x_{i}^N-y_{N-j+1},t))_{1\leq i,j \leq N},
\end{equation}
where
\begin{equation}
{\cal D'}=\{x_i^n,2 \leq i \leq n \leq N | x_i^n\geq x_{i-1}^{n-1}\}.
\end{equation}
Then, using the antisymmetry of the determinant and Lemma~\ref{Lem2} below we can rewrite (\ref{eqLem2}) as
\begin{equation}
G(x_1,\ldots,x_N;t)=\sum_{\cal D}
\det(F_{-j+1}(x_{i}^N-y_{N-j+1},t))_{1\leq i,j \leq N},
\end{equation}
where
\begin{equation}
{\cal D}=\{x_i^n,2 \leq i \leq n \leq N | x_i^n>x_i^{n+1},x_i^n\geq x_{i-1}^{n-1}\}.
\end{equation}

\begin{lem}\label{Lem2} Let $f$ an antisymmetric function of $\{x_1^{N},\ldots,x_N^{N}\}$. Then, whenever $f$ has enough decay to make the sums finite,
\begin{equation}
\sum_{\cal D}f(x_1^{N},\ldots,x_N^{N})=\sum_{\cal D'}f(x_1^{N},\ldots,x_N^{N}),
\end{equation}
with the positions $x_1^1>x_1^2>\ldots>x_1^N$ being fixed.
\end{lem}

Now, notice that, for $n=-k<0$, (\ref{eqFn}) has only a pole at $w=0$, which implies that
\begin{equation}
F_{n+1}(x,t)=-\sum_{y<x}F_n(x,t).
\end{equation}
Define then
\begin{equation}\label{eq58}
\begin{aligned}
\Psi^N_{k}(x)&:=\frac{1}{2\pi \I} \oint_{\Gamma_{0}} \frac{\dx w}{w}\frac{(1-w)^{k}}{w^{x-y_{N-k}-k}}e^{t(w-1)},\\
\phi(x,y)&:=\Id(x>y),\quad \phi({\rm virt},y)=1.
\end{aligned}
\end{equation}
In particular, for $k=1,\ldots,N$, $\Psi^N_{k}(x)=(-1)^{k}F_{-k}(x-y_{N-k},t)$.

\begin{figure}
\begin{center}
  \psfrag{m}[c]{$<$} \psfrag{e}[c]{$\leq$} \psfrag{x11}{$x_1^1$}
  \psfrag{x12}{$x_1^2$} \psfrag{x13}{$x_1^3$} \psfrag{x14}{$x_1^4$}
  \psfrag{x22}{$x_2^2$} \psfrag{x23}{$x_2^3$} \psfrag{x24}{$x_2^4$}
  \psfrag{x33}{$x_3^3$} \psfrag{x34}{$x_3^4$} \psfrag{x44}{$x_4^4$}
  \includegraphics[height=3.5cm]{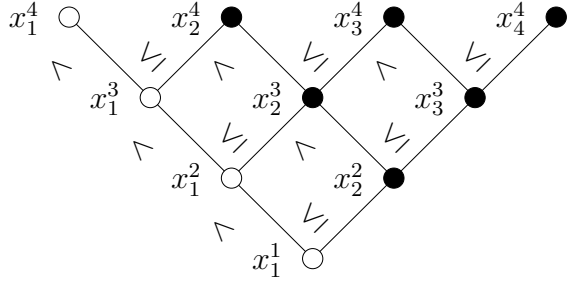}
\caption{Graphical representation of the domain of integration $\cal D$ for $N=4$. One has to ``integrate'' out the variables $x_i^j$, $i\geq 2$ (i.e., the black dots). The positions of $x_1^k$, $k=1,\ldots,N$ are given (i.e., the white dots).}
\label{figScheme}
\end{center}
\end{figure}

For a moment consider particle configurations ordered level-by-level, i.e., with $x_1^n\leq x_2^n\leq \ldots\leq x_n^n$ for all $n=1,\ldots,N$. Then, the \textbf{interlacing condition} $\cal D$, is given by
\begin{equation}\label{eqInterlacingTASEP}
x_1^{n+1}<x_1^{n}\leq x_2^{n+1} < x_2^{n}\leq \ldots<x_{n}^{n} \leq x_{n+1}^{n+1},
\end{equation}
for $n=1,\ldots,N-1$ (see Figure~\ref{figScheme} for a graphical representation).
This interlacing structure can be written as
\begin{equation}
\det[\phi(x_i^{n},x_j^{n+1})]_{1\leq i,j\leq n+1}=\left\{
\begin{array}{ll}
1,&\textrm{if (\ref{eqInterlacingTASEP}) is satisfied},\\
0,&\textrm{otherwise},
\end{array}
\right.
\end{equation}
where $x_{n+1}^n={\rm virt}$. Therefore, we can replace the sum over $\cal D$ by a product of determinants of increasing sizes. Namely,
\begin{equation}
G(x_1,\ldots,x_N;t) = \sum_{\begin{subarray}{c}
x_k^n\in\Z\\2\leq k\leq n\leq N
\end{subarray}} Q(\{x_k^n,1\leq k\leq n\leq N\})
\end{equation}
where the measure $Q$ is
\begin{equation}\label{eqProdDets}
\begin{aligned}
Q(\{x_k^n,1\leq k\leq n\leq N\}) &=
\Big(\prod_{n=1}^{N-1}
\det(\phi(x_i^{n},x_j^{n+1}))_{1\leq i,j \leq n+1}\Big) \\
&\times \det(\Psi^N_{N-j}(x^N_{i}))_{1\leq i,j \leq N}.
\end{aligned}
\end{equation}
As for the GUE minor case, if we compute $\Psi^{n}_{n-k}(x):=(\phi * \Psi^{n+1}_{n+1-k})(x)$, we get the (\ref{eq58}) but with $N$ replaced by $n$. Notice that (\ref{eqProdDets}) is symmetric under exchange of variables at the same level, so that we can relax the constraint that the particle configurations are ordered level-by-level, by just multiplying (\ref{eqProdDets}) by a constant.

\begin{remark}
The measure (\ref{eqProdDets}) is not necessarily a probability measure, since positivity is not ensured, but still has the same structure as the GUE minors measure, compare with (\ref{eqMeasureGUEminors}). In particular, the distribution of TASEP particles' position can be expressed in the same form as the gap probability, but in general for a signed determinantal point process. The precise statement is in Theorem~\ref{ThmJointCorr} below.
\end{remark}

Applying the discrete version of Theorem~\ref{ThmSasa} (in which one just replaces $\R$ by $\Z$ and integrals by sums), see Lemma~3.4 of~\cite{BFPS06}, we get the following result.
\begin{thm}\label{ThmJointCorr}
Let us start at time $t=0$ with $N$ particles at positions \mbox{$y_N < \ldots < y_2 < y_1$}.  Let $\sigma(1)<\sigma(2)<\ldots<\sigma(m)$ be the indices of $m$ out of the $N$ particles. The joint distribution of their positions $x_{\sigma(k)}(t)$ is given by
\begin{equation}
\Pb\Big(\bigcap_{k=1}^m \big\{x_{\sigma(k)}(t) \geq s_k\big\}\Big)=
\det(\Id-\chi_s K_t\chi_s)_{\ell^2(\{\sigma(1),\ldots,\sigma(m)\}\times\Z)}
\end{equation}
where $\chi_s(\sigma(k),x)=\Id(x<s_k)$. $K_t$ is the extended kernel with entries
\begin{equation}\label{eqKernelFinal}
K_t(n_1,x_1;n_2,x_2)=-\phi^{(n_1,n_2)}(x_1,x_2)
+\sum_{k=1}^{n_2} \Psi^{n_1}_{n_1-k}(x_1) \Phi^{n_2}_{n_2-k}(x_2)
\end{equation}
where
\begin{equation}
\phi^{(n_1,n_2)}(x_1,x_2) = \binom{x_1-x_2-1}{n_2-n_1-1},
\end{equation}
\begin{equation}\label{eq2.4}
\Psi_i^{n}(x)=\frac{1}{2\pi \I} \oint_{\Gamma_0} \frac{\dx w}{w^{i+1}}
\frac{(1-w)^{i}}{w^{x-y_{n-i}}}e^{t(w-1)},
\end{equation}
and the functions $\Phi_i^{n}(x)$, $i=0,\ldots,n-1$, form a family
of polynomials of degree $\le n$ satisfying
\begin{equation}
\sum_{x\in\Z}\Psi_i^{n}(x)\Phi_j^{n}(x)=\delta_{i,j}.
\label{ortho}
\end{equation}
The path $\Gamma_0$ in the definition of $\Psi_i^{n}$ is any simple
loop, anticlockwise oriented, which includes the pole at $w=0$ but
\emph{not} the one at $w=1$.
\end{thm}
Remark that the initial particle positions, $y_1,\ldots,y_N$, are in the definition of the $\Psi_i^n$'s and consequently enters in the $\Phi_i^n$'s through the orthogonality relation (\ref{ortho}).

\subsection{Correlation functions for step initial conditions: Charlier polynomials}\label{subsectTASEPCharlier}
Now we consider the particular case of step initial conditions, $y_k=-k$ for $k\geq 1$. In that case, the measure (\ref{eqProdDets}) is positive, i.e., it is a probability measure. The correlation functions of subsets of $\{x_k^n(t),1\leq k\leq n,n\geq 1\}$ are determinantal with kernel $K_t^{\rm TASEP}$, which is computed by Theorem~\ref{ThmJointCorr}.

The correlation kernel $K_t^{\rm TASEP}$ is given in terms of Charlier polynomials, which we now introduce. The Charlier polynomial of degree $n$ is denoted by $C_n(x,t)$ and defined as follows. Consider the weight $\omega$ on $\Z_+=\{0,1,\ldots\}$
\begin{equation}
\omega(x)=e^{-t} t^x/x!.
\end{equation}
Then, $C_n(x,t)$ is defined via the orthogonal condition
\begin{equation}
\sum_{x\geq 0} \omega(x) C_n(x,t) C_m(x,t)=\frac{n!}{t^n}\delta_{n,m}
\end{equation}
or, equivalently, $C_n(x,t)=(-1/t)^n x^n+\cdots$. They can be expressed in terms of hypergeometric functions
\begin{equation}\label{eqCh}
C_n(x,t)=\phantom{}_2F_0(-n,-x;\, ;-1/t) = C_x(n,t).
\end{equation}
From the generating function of the Charlier polynomials
\begin{equation}
\sum_{n\geq 0}\frac{C_n(x,t)}{n!}(tw)^n=e^{wt}(1-w)^x
\end{equation}
one gets the integral representation
\begin{equation}\label{eq62}
\frac{t^n}{n!} C_n(x,t)=\frac{1}{2\pi \I}\oint_{\Gamma_0}\dx w
\frac{e^{wt}(1-w)^x}{w^{n+1}}.
\end{equation}
Equations (\ref{eqCh}) and (\ref{eq62}) then give
\begin{equation}
\Psi^n_{k}(-n+x)=\frac{e^{-t} t^{x}}{x!} C_{k}(x,t)
\quad \textrm{and} \quad \Phi^n_{j}(-n+x)=\frac{t^j}{j!} C_j(x,t).
\end{equation}
In particular, the kernel of the joint distributions of $\{x_1^N,\ldots,x_N^N\}$ is
\begin{equation}\label{eq73}
\begin{aligned}
K^{\rm TASEP}_t(N,-N+x;N,-N+y)&=\sum_{k=0}^{N-1}\Psi^N_k(-N+x)\Phi^N_k(-N+y)\\
&=\omega(x) \sum_{k=0}^{N-1} q_k(x) q_k(y)
\end{aligned}
\end{equation}
with $q_k(x)=(-1)^k\frac{t^{k/2}}{\sqrt{k!}} C_k(x,t)=(t^{k} k!)^{-1/2} x^k+\ldots$, orthogonal polynomials with respect to $\omega(x)$. Therefore, using the Christoffel-Darboux formula (\ref{eqChristDarb}) we get, for $x\neq y$,
\begin{multline}
K^{\rm TASEP}_t(N,-N+x;N,-N+y)\\ = \frac{e^{-t} t^{x}}{x!} \frac{t^N}{(N-1)!}\frac{C_{N-1}(x,t)C_N(y,t)-C_N(x,t)C_{N-1}(y,t)}{x-y}.
\end{multline}

\begin{remark}
The expression of the kernel in (\ref{eq73}) is not symmetric in $x$ and $y$. Is this wrong? No! For a determinantal point process the kernel is not uniquely defined, but only up to conjugation. For any function $f>0$, the kernel $\widetilde K(x,y)=f(x) K(x,y) f(y)^{-1}$ describes the same determinantal point process of the kernel $K(x,y)$. Indeed, the $f$ cancels out in the determinants defining the correlation functions. In particular, by choosing $f(x)=1/\sqrt{\omega(x)}$, we get a symmetric version of (\ref{eq73}).
\end{remark}

\subsubsection*{Double integral representation of $K_t^{\rm TASEP}$}
Another typical way of representing the kernel $K_t^{\rm TASEP}$ is as double integral. This representation is well adapted to large-$t$ asymptotic analysis (both in the bulk and the edge).

Let us start with (\ref{eq2.4}) with $y_k=-k$. We have
\begin{equation}
\Psi^n_k(x)=\frac{1}{2\pi\I}\oint_{\Gamma_0}\dx w \frac{e^{t(w-1)}(1-w)^k}{w^{x+n+1}},
\end{equation}
and remark that $\Psi^n_k(x)=0$ for $x<-n$, $k\geq 0$. The orthogonal functions $\Phi^n_k(x)$, $k=0,\ldots,n-1$, have to be polynomials of degree $k$ and to satisfy the orthogonal relation $\sum_{x\geq -n} \Psi^n_k(x)\Phi^n_j(x)=\delta_{i,j}$. They are given by
\begin{equation}
\Phi^n_j(x)=\frac{-1}{2\pi\I}\oint_{\Gamma_1}\dx z \frac{e^{-t(z-1)} z^{x+n}}{(1-z)^{j+1}}.
\end{equation}
Indeed, for any choice of the paths $\Gamma_0, \Gamma_1$ such that $|z|<|w|$,
\begin{equation}
\begin{aligned}
\sum_{x\geq -n} \Psi^n_k(x)\Phi^n_j(x) &= \frac{-1}{(2\pi\I)^2}\oint_{\Gamma_1}\dx z \oint_{\Gamma_{0,z}} \dx w \frac{e^{t(w-1)}}{e^{t(z-1)}}\frac{(1-w)^k}{(1-z)^{j+1}}\sum_{x\geq -n}\frac{z^{x+n}}{w^{x+n+1}} \\
&=\frac{-1}{(2\pi\I)^2}\oint_{\Gamma_1}\dx z \oint_{\Gamma_{0,z}} \dx w \frac{e^{t(w-1)}}{e^{t(z-1)}}\frac{(1-w)^k}{(1-z)^{j+1}}\frac{1}{w-z} \\
&=\frac{(-1)^{k-j}}{2\pi\I}\oint_{\Gamma_1}\dx z (z-1)^{k-j-1}=\delta_{j,k},
\end{aligned}
\end{equation}
since the only pole in the last $w$-integral is the simple pole at $w=z$.

After the orthogonalization, we can determine the kernel. Let us compute the last term in (\ref{eqKernelFinal}). For $k>n_2$, we have $\Phi^{n_2}_{n_2-k}(x)=0$. So, by choosing $z$ close enough to $1$, we have $|1-z|<|1-w|$, and we can extend the sum over $k$ to $+\infty$, i.e.,
\begin{equation}\label{eq76}
\begin{aligned}
\sum_{k=1}^{n_2}\Psi^{n_1}_{n_1-k}(x_1)\Phi^{n_2}_{n_2-k}(x_2)&= \sum_{k=1}^{n_2} \frac{-1}{(2\pi\I)^2} \oint_{\Gamma_0}\dx w \oint_{\Gamma_1}\dx z \frac{e^{tw}z^{x_2+n_2}}{e^{tz}w^{x_1+n_1+1}} \frac{(1-w)^{n_1-k}}{(1-z)^{n_2-k+1}}\\
&=\frac{-1}{(2\pi\I)^2} \oint_{\Gamma_0}\dx w\oint_{\Gamma_1}\dx z  \frac{e^{tw}z^{x_2+n_2}}{e^{tz}w^{x_1+n_1+1}} \sum_{k=1}^{n_2} \frac{(1-w)^{n_1-k}}{(1-z)^{n_2-k+1}} \\
&=\frac{-1}{(2\pi\I)^2} \oint_{\Gamma_0}\dx w\oint_{\Gamma_1}\dx z  \frac{e^{tw}z^{x_2+n_2}}{e^{tz}w^{x_1+n_1+1}} \sum_{k=1}^{\infty} \frac{(1-w)^{n_1-k}}{(1-z)^{n_2-k+1}} \\
&=\frac{-1}{(2\pi\I)^2} \oint_{\Gamma_0}\dx w \oint_{\Gamma_1}\dx z \frac{e^{tw}(1-w)^{n_1} z^{x_2+n_2}}{e^{tz}(1-z)^{n_2}w^{x_1+n_1+1}} \frac{1}{z-w}.
\end{aligned}
\end{equation}
The integral over $\Gamma_1$ contains only the pole $z=1$, while the integral over $\Gamma_0$ only the pole $w=0$ (i.e., $w=z$ is not inside the integration paths). This is the kernel for $n_1\geq n_2$. For $n_1<n_2$, there is the extra term $-\phi^{(n_1,n_2)}(x_1,x_2)$. It is not difficult to check that
\begin{equation}
\begin{aligned}
&\frac{1}{(2\pi\I)^2} \oint_{\Gamma_0}\dx w \oint_{\Gamma_w}\dx z\frac{e^{tw}(1-w)^{n_1} z^{x_2+n_2}}{e^{tz}(1-z)^{n_2}w^{x_1+n_1+1}} \frac{1}{z-w}\\
&=\frac{1}{2\pi\I} \oint_{\Gamma_0}\dx w \frac{(1-w)^{n_1-n_2}}{w^{x_1+n_1-(x_2+n_2)+1}}=\binom{x_1-x_2-1}{n_2-n_1-1}.
\end{aligned}
\end{equation}
Therefore, the double integral representation of $K_t^{\rm TASEP}$ (for step initial condition) is the following:
\begin{equation}\label{eqKernelIntegrals}
\begin{aligned}
&K_t^{\rm TASEP}(n_1,x_1;n_2,x_2) \\
&=\left\{
\begin{array}{ll}
{\displaystyle \frac{-1}{(2\pi\I)^2} \oint_{\Gamma_0}\dx w \oint_{\Gamma_1}\dx z \frac{e^{tw}(1-w)^{n_1} z^{x_2+n_2}}{e^{tz}(1-z)^{n_2}w^{x_1+n_1+1}} \frac{1}{z-w}}, &\textrm{ if }n_1\geq n_2,\\[1em]
{\displaystyle \frac{-1}{(2\pi\I)^2} \oint_{\Gamma_0}\dx w \oint_{\Gamma_{1,w}}\dx z \frac{e^{tw}(1-w)^{n_1} z^{x_2+n_2}}{e^{tz}(1-z)^{n_2}w^{x_1+n_1+1}} \frac{1}{z-w}}, &\textrm{ if }n_1<n_2.
\end{array}
\right.
\end{aligned}
\end{equation}

\begin{remark}
Notice that in the GUE case, the most natural object are the eigenvalues for the $N\times N$ matrix, $\lambda_1^N,\ldots,\lambda_N^N$. They are directly associated with a determinantal point process. The corresponding quantities in terms of TASEP are not the positions of the particles, $x_1^1,\ldots,x_1^N$. The measure on these particle positions is not determinantal, but with the extension to the larger picture, namely to $\{x_k^n,1\leq k\leq n\leq N\}$, we recover a determinantal structure. This is used to determine the joint distributions of our particles, since in terms of $\{x_k^n,1\leq k\leq n\leq N\}$ we need only to compute a gap probability.
\end{remark}

\subsection{Discrete time TASEP}\label{subsectTASEPdiscrete}
There are several discrete time dynamics of TASEP from which the continuous time limit can be obtained. The most common dynamics are:
\begin{itemize}
  \item \textbf{Parallel update}: at time $t\in \Z$ one first selects which particles can jump (their right neighboring site is empty). Then, the configuration of particles at time $t+1$ is obtained by moving independently with probability $p\in (0,1)$ the selected particles.
  \item \textbf{Sequential update}: one updates the particles from right to left. The configuration at time $t+1$ is obtained by moving with probability $p\in (0,1)$ the particles whose right rite is empty. This procedure is from right to left, which implies that also a block of $m$ particles can move in one time-step with probability $p^m$.
\end{itemize}
Other dynamical rules have also been introduced, see~\cite{Sch00} for a review.

\subsubsection*{Sequential update}
For the TASEP with sequential update, there is an analogue of Lemma~\ref{lem1}, with the only difference lying in the functions $F_n$ (see~\cite{RS05} and Lemma~3.1 of~\cite{BFP06}). The functions $F_n$ satisfy again the recursion relation (\ref{eqRecRel}) and Theorem~\ref{ThmJointCorr} still holds with the only difference being in the $\Psi^n_i(x)$'s which are now given by
\begin{equation}
\Psi^n_i(x)=\frac{1}{2\pi\I}\oint_{\Gamma_0}\frac{\dx w}{w^{i+1}} \frac{(1-p+pw)^t (1-w)^i}{w^{x-y_{n-i}}}.
\end{equation}
For step initial conditions, the kernel is then given by (\ref{eqKernelIntegrals}) with $e^{tw}/e^{tz}$ replaced by $(1-p+pw)^t/(1-p+pz)^t$.

\subsubsection*{Parallel update}
For the TASEP with parallel update, the same formalism used above can still be applied. However, the interlacing condition and the transition functions $\phi_n$ are different. The details can be found in Section~3 of~\cite{BFS07b} (in that paper we also consider the case of different times, which can be used for example to study the tagged particle problem). The analogue of (\ref{eqProdDets}) is the following (see Proposition 7 of~\cite{BFS07b}).
\begin{lem}
The transition probability $G(x;t)$ can be written as a sum over
\begin{equation}
{\cal D}''=\{x_i^n,2\leq i\leq n\leq N | x_i^n>x_{i-1}^n\}
\end{equation}
as follows:
\begin{equation}
G(x_1,\ldots,x_N;t)=\sum_{{\cal D}''} \widetilde Q(\{x_k^n,1\leq k\leq n\leq N\}),
\end{equation}
where
\begin{equation}\label{eqProdDetsParallel}
\begin{aligned}
\widetilde Q(\{x_k^n,1\leq k\leq n\leq N\}) &=
\Big(\prod_{n=1}^{N-1}
\det(\phi^\sharp(x_{i-1}^{n},x_j^{n+1}))_{1\leq i,j \leq n+1}\Big) \\
&\times \det(F_{-j+1}(x^N_{i}-y_{N-j+1},t+1-j))_{1\leq i,j \leq N}.
\end{aligned}
\end{equation}
where we set $x_{0}^{n}=-\infty$ (we call them virtual variables). The function $\phi^\sharp$ is defined by
\begin{equation}
\phi^\sharp(x,y)=\left\{
\begin{array}{ll}
1,&y\geq x,\\ p,& y=x-1 \\ 0, & y\leq x-2,
\end{array}
\right.
\end{equation}
and $F_{-n}$ is given by
\begin{equation}
F_{-n}(x,t)=\frac{1}{2\pi\I}\oint_{\Gamma_{0,-1}}\dx w \frac{w^n}{(1+w)^{x+n+1}}(1+pw)^t.
\end{equation}
\end{lem}
The product of determinants in (\ref{eqProdDetsParallel}) also implies a \textbf{weighted interlacing condition}, different from the one of TASEP. More precisely,
\begin{equation}
x_i^{n+1}\leq x_i^n-1 \leq x_{i+1}^{n+1}.
\end{equation}
The difference between the continuous-time TASEP is that it can happens that $x_{i+1}^{n+1}=x_i^n-1$. However, for each occurrence of such a configuration, the weight is multiplied by $p$. The continuous-time limit is obtained by replacing $t$ by $t/p$ and letting $p\to 0$.

Also in this case Theorem~\ref{ThmJointCorr} holds with the following new functions,
\begin{equation}
\begin{aligned}
\widetilde \phi^{(n_1,n_2)}(x_1,x_2)&=\frac{1}{2\pi\I}\oint_{\Gamma_{0,-1}} \hspace{-1em} \dx w \frac{1}{(1+w)^{x_1-x_2+1}}
\left(\frac{w}{(1+w)(1+pw)}\right)^{n_1-n_2} \hspace{-1em} \Id_{[n_2>n_1]},\\
\Psi^n_{i}(x)&=\frac{1}{2\pi\I}\oint_{\Gamma_{0,-1}} \hspace{-1em}\dx w \frac{(1+pw)^{t}}{(1+w)^{x-y_{n-i}+1}}
\left(\frac{w}{(1+w)(1+pw)}\right)^{i},
\end{aligned}
\end{equation}
and $\Phi^n_k$ are polynomials of degree $k$ given by the orthogonality condition (\ref{ortho}): $\sum_{x\in\Z}\Psi_i^{n}(x)\Phi_j^{n}(x)=\delta_{i,j}$. In particular, for step initial conditions, $y_k=-k$, $k\geq 1$, we obtain the following result.
\begin{prop}
The correlation kernel for discrete-time TASEP with parallel update is given by
\begin{equation}
K_t^{\rm TASEP}(n_1,x_1;n_2,x_2)=-\phi^{(n_1,n_2)}(x_1,x_2) + \widetilde K_t^{\rm TASEP}(n_1,x_1;n_2,x_2),
\end{equation}
with $\phi^{(n_1,n_2)}$ given by
\begin{equation}
\phi^{(n_1,n_2)}(x_1,x_2)=\frac{\Id_{[n_2>n_1]}}{2\pi\I}\oint_{\Gamma_{-1}}\dx w \frac{(1+pw)^{n_2-n_1}w^{n_1-n_2} }{(1+w)^{(x_1+n_1)-(x_2+n_2)+1}}
\end{equation}
and
\begin{multline}
\widetilde K_t^{\rm TASEP}(n_1,x_1;n_2,x_2) \\ =\frac{1}{(2\pi\I)^2}\oint_{\Gamma_0}\dx z \oint_{\Gamma_{-1}}\dx z \frac{(1+pw)^{t-n_1+1}}{(1+pz)^{t-n_2+1}}\frac{w^{n_1}(1+z)^{x_2+n_2}}{z^{n_2}(1+w)^{x_1+n_1+1}}\frac{1}{w-z}.
\end{multline}
\end{prop}

\begin{remark}
The discrete-time parallel update TASEP with step initial condition is equivalent to the shuffling algorithm on the Aztec dynamics as shown in~\cite{Nor08}. This particle dynamics also fits in the framework developed in~\cite{BF08}. See~\cite{FerAZTEC} for an animation, where the particles have coordinates \mbox{$(z_i^n:=x_i^n+n,n)$}.
\end{remark}

\section{$2+1$ dynamics: connection to random tilings and random matrices}\label{sect2plus1growth}
In recent years there has been a lot of progress in understanding large time fluctuations of driven interacting particle systems on the one-dimensional lattice. Evolution of such systems is commonly interpreted as random growth of a one-dimensional interface, and if one views the time as an extra variable, the evolution produces a random surface (see e.g.~Figure 4.5 in~\cite{Pra03} for a nice illustration). In a different direction, substantial progress have also been achieved in studying the asymptotics of random surfaces arising from dimers on planar bipartite graphs.

Although random surfaces of these two kinds were shown to share certain asymptotic properties (also common to random matrix models), no direct connection between them was known.  We present a class of two-dimensional random growth models (that is, the principal object is a randomly growing surface, embedded in the four-dimensional space-time).

In two different projections these models yield random surfaces of the two kinds mentioned above (one reduces the spatial dimension by one, the second projection is to fixed time).

Let us now we explain the $2+1$-dimensional dynamics. Consider the set of variables $\{x_k^n(t),1\leq k\leq n, n\geq 1\}$ and let us see what is their evolution inherited from the TASEP dynamics.

\subsection{$2+1$ dynamics for continuous time TASEP}\label{subsectTASEPandGrowth}
\subsubsection*{Packed initial condition}
\begin{figure}[t!]
\begin{center}
\psfrag{x}[b]{$x$}
\psfrag{n}[b]{$n$}
\psfrag{h}[b]{$h$}
\psfrag{x11}[cl]{$x_1^1$}
\psfrag{x12}[cl]{$x_1^2$}
\psfrag{x13}[cl]{$x_1^3$}
\psfrag{x14}[cl]{$x_1^4$}
\psfrag{x15}[cl]{$x_1^5$}
\psfrag{x22}[cl]{$x_2^2$}
\psfrag{x23}[cl]{$x_2^3$}
\psfrag{x24}[cl]{$x_2^4$}
\psfrag{x25}[cl]{$x_2^5$}
\psfrag{x33}[cl]{$x_3^3$}
\psfrag{x34}[cl]{$x_3^4$}
\psfrag{x35}[cl]{$x_3^5$}
\psfrag{x44}[cl]{$x_4^4$}
\psfrag{x45}[cl]{$x_4^5$}
\psfrag{x55}[cl]{$x_5^5$}
\psfrag{tasep}[c]{\textbf{TASEP}}
\includegraphics[width=\textwidth]{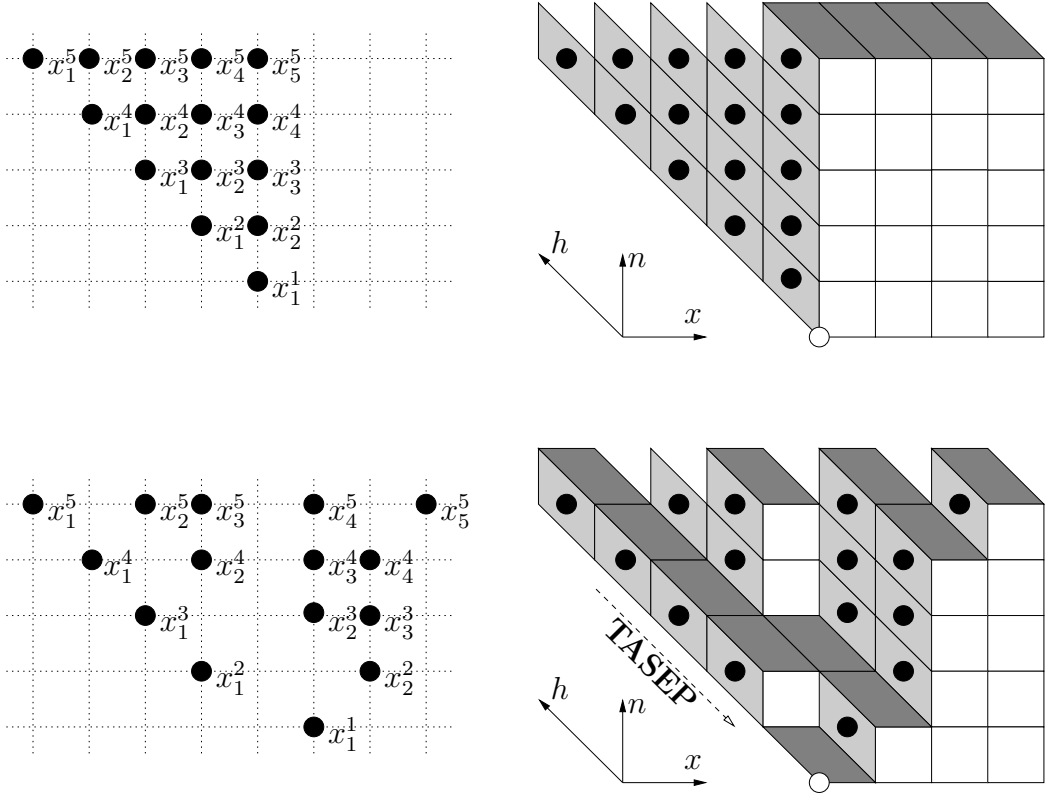}
\caption{(Top, left) Illustration of the initial conditions for the particles system. (Bottom, left) A configuration obtained from the initial conditions. (right) The corresponding lozenge tiling configurations. In the height function picture, the white circle has coordinates $(x,n,h)=(-1/2,0,0)$. For a Java animation of the model see~\cite{FerAKPZ}.}
\label{Figure2dDynamics}
\end{center}
\end{figure}

Consider continuous time TASEP with step-initial condition, $y_k=-k$ for \mbox{$k\geq 1$}. As we will verify below, a further property of the measure (\ref{eqProdDets}) with step initial conditions is that
\begin{equation}\label{eq61}
x_k^n(0)=-n+k-1,
\end{equation}
i.e., step initial condition for TASEP induces naturally a packed initial condition for the $2+1$ dynamics, which is illustrated in Figure~\ref{Figure2dDynamics} (top left picture).

Let us verify that (\ref{eq61}) holds. The first $N-1$ determinants in (\ref{eqProdDets}) imply the interlacing condition (\ref{eqInterlacingTASEP}). In particular, $x_1^N(0)=-N$ and \mbox{$x_k^N(0)\geq -N+k-1$} for $k\geq 2$.
At time $t=0$ we have
\begin{equation}
F_{-k}(x,0)=\frac{1}{2\pi\I}\oint_{\Gamma_0}\dx w \frac{(w-1)^k}{w^{x+k+1}}.
\end{equation}
By Cauchy residue's theorem, we have $F_{-k}(x,0)=0$ for $x\geq 1$ (since the integrand has no pole at $\infty$), $F_{-k}(x,0)=0$ for $x<-k$ (no pole at $0$), and $F_{-k}(0,0)=1$. The last determinant in (\ref{eqProdDets}) is then the determinant of
\begin{equation}\label{eq62B}
\left[
  \begin{array}{cccc}
    F_0(0,0) & F_{-1}(-1,0) & \cdots & F_{-N+1}(-N+1,0) \\
    F_0(x_2^N(0)+N,0) & F_{-1}(x_2^N(0)+N-1) & \cdots & F_{-N+1}(x_2^N(0)+1,0) \\
    \vdots & \vdots & \ddots & \vdots \\
    F_0(x_N^N(0)+N,0) & F_{-1}(x_N^N(0)+N-1,0) & \cdots & F_{-N+1}(x_N^N(0)+1,0), \\
  \end{array}
\right].
\end{equation}
Let us determine when the determinant of the matrix (\ref{eq62B}) is nonzero:
\begin{enumerate}
\item Because of $x_k^N(0)\geq -N+k-1$, the first column of (\ref{eq62B}) is $[1,0,\ldots,0]^t$.
\item Then, if $x_2^N(0)>-N+1$, the second column is $[*,0,\ldots,0]^t$ and $\det(\ref{eq62B})=0$. Thus we have $x_2^N(0)=-N+1$.
\item Repeating the argument for the other columns, we obtain that $\det(\ref{eq62B})\neq 0$ if and only if $x_k^N(0)=-N+k-1$ for $k=3,\ldots,N$.
\end{enumerate}

This initial condition is illustrated in Figure~\ref{Figure2dDynamics} (top, left).

\subsubsection*{Dynamics}
Now we explain the dynamics on the variables $\{x_k^n(t),1\leq k\leq n,n\geq 1\}$ which is inherited by the dynamics on the TASEP particles $\{x_1^n(t),n\geq 1\}$. Each of the particles $x_k^m$ has an independent exponential clock of rate one, and when the $x_k^m$-clock rings the particle attempts to jump to the right by one. If at that moment $x_k^m=x_k^{m-1}-1$ then the jump is blocked. If that is not the case, we find the largest $c\geq 1$ such that
$x_k^m=x_{k+1}^{m+1}=\dots=x_{k+c-1}^{m+c-1}$, and all $c$ particles in this string jump to the right by one.

Informally speaking, the particles with smaller upper indices are heavier than those with larger upper indices, so that the heavier particles block and push the lighter ones in order for the interlacing conditions to be preserved.

We illustrate the dynamics using Figure~\ref{Figure2dDynamics}, which shows a possible configuration of particles obtained from our initial condition. If in this state of the system the $x_1^3$-clock rings, then particle $x_1^3$ does not move, because it is blocked by particle $x_1^2$. If it is the $x_2^2$-clock that rings, then particle $x_2^2$ moves to the right by one unit, but to keep the interlacing property satisfied, also particles $x_3^3$ and $x_4^4$ move by one unit at the same time. This aspect of the dynamics is called ``pushing''.

\subsection{Interface growth interpretation}\label{subsectGrowhInterpr}
Figure~\ref{Figure2dDynamics} (right) has a clear three-dimensional connotation. Given the random configuration $\{x_k^n(t)\}$ at time moment $t$, define the random {\it height function}
\begin{equation}\label{eqDefinHeight}
\begin{aligned}
&h:(\Z+\tfrac 12)\times \Z_{>0}\times \R_{\geq 0}\to \Z_{\geq 0},\\
&h(x,n,t)=\#\{k\in\{1,\dots,n\}\mid x_k^n(t)> x\}.
\end{aligned}
\end{equation}
In terms of the tiling on Figure~\ref{Figure2dDynamics}, the height function is defined at the vertices of rhombi, and it counts the number of particles to the right from a given vertex. (This definition differs by a simple linear function of $(x,n)$ from the standard definition of the height function for lozenge tilings, see e.g.~\cite{Ken04,KenLectures}.) The initial condition corresponds to starting with perfectly flat facets.

In terms of the step surface of Figure~\ref{Figure2dDynamics}, the evolution consists of removing all columns of $(x,n,h)$-dimensions $(1,*,1)$ that could be removed, independently with exponential waiting times of mean one. For example, if $x_2^2$ jumps to its right, then three consecutive cubes
(associated to $x_2^2,x_3^3,x_4^4$) are removed. Clearly, in this dynamics the directions $x$ and $n$ do not play symmetric roles. Indeed, this model belongs to the $2+1$ anisotropic KPZ class of stochastic growth models, see~\cite{BF08,BF08b}.

\subsection{Random tilings interpretation}\label{subsectRandomTiling}
A further interpretation of our particles' system is a random tiling model. To see that one surrounds each particle location by a rhombus of one type (the light-gray in Figure~\ref{Figure2dDynamics}) and draws unit-length horizontal edges through locations where there are no particles. In this way we have a random tiling with three types of tiles that we call \emph{white}, \emph{light-gray}, and \emph{dark-gray}. Our initial condition corresponds to a perfectly regular tiling.

\begin{figure}
\begin{center}
\psfrag{rate}[l]{rate $1$}
\includegraphics[height=5cm]{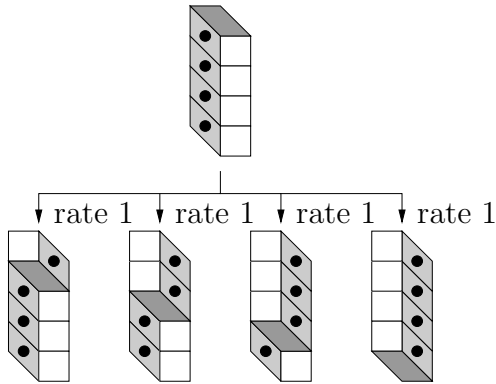}
\caption{Illustration of the dynamics on tiles for a column of height $m=4$.}
\label{FigTiling}
\end{center}
\end{figure}

The dynamics on random tilings is the following. Consider all sub-configuration of the random tiling which looks like a visible column, i.e., for some $m\geq 1$, there are $m$ light-gray tiles on the left of $m$ white tiles (and then automatically closed by a dark-gray tile). The dynamics is an exchange of light-gray and white tiles within the column. More precisely, for a column of height $m$, for all $k=1,\ldots,m$, independently and with rate $1$, there is an exchange between the top $k$ light-gray tiles with the top white tiles as illustrated in Figure~\ref{FigTiling} for the case $m=4$.

\begin{remark}
We can also derive a determinantal formula not only for the correlation of light-gray tiles, but also for the three types of tiles. This is explicitly stated in Theorem 5.2 of~\cite{BF08}.
\end{remark}

\subsection{Diffusion scaling and relation with GUE minors}\label{subsectGUEscalingLimit}
There is an interesting \emph{partial} link with GUE minors. In the diffusion scaling limit
\begin{equation}
\xi_k^n:=\sqrt{2N}\lim_{t\to\infty}\frac{x_k^n(t)-t}{\sqrt{2t}}
\end{equation}
the measure on $\{\xi_k^n,1\leq k\leq n \leq N\}$ is exactly given by (\ref{eqMeasureGUEminors}).
\begin{remark}
It is important to stress, that this correspondence is a fixed-time result. From this, a dynamical equivalence does not follow. Indeed, if we let the GUE matrices evolves according to the so-called Dyson's Brownian Motion dynamics, then the evolution of the minors is not the same as the (properly rescaled) evolution from our $2+1$ dynamics for TASEP~\cite{ANvM10b}. Nevertheless, projecting onto the $(t,n)$ paths with increasing $t$ and decreasing $n$ one still obtains the same measures~\cite{FF10}.
\end{remark}

\subsection{Shuffling algorithm and discrete time TASEP}\label{subsectAztec}
\begin{figure}
\begin{center}
\includegraphics[height=5cm]{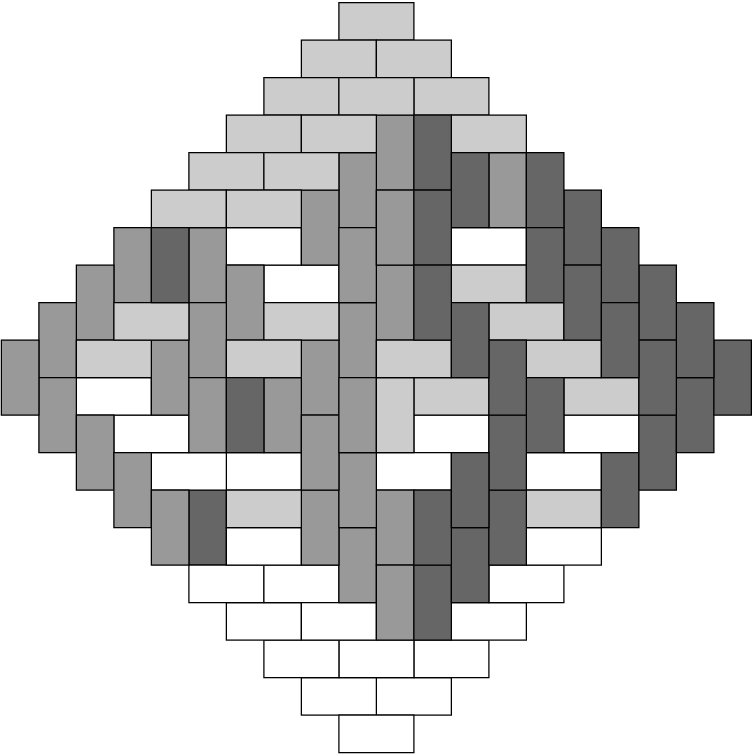}
\caption{A random tiling of the Aztec diamond of size $n=10$.}
\label{FigAztecDiamond}
\end{center}
\end{figure}

An Aztec diamond is a shape like the outer border of Figure~\ref{FigAztecDiamond}. The shuffling algorithm~\cite{EKLP92,JPS98} provides a way of generating after $n$ steps uniform sample of an Aztec diamond of size $n$ tiled with dominos.

We now discuss (not prove) the connection between discrete time TASEP with parallel update and step initial condition. Moreover, we take the parameter $p=1/2$ to get uniform distribution. It is helpful to do a linear change of variable. Instead of $x_k^n$ we use
\begin{equation}
z_k^n=x_k^n+n,
\end{equation}
so that the interlacing condition becomes
\begin{equation}\label{eqInterAztec}
z_k^{n+1}\leq z_k^n\leq z_{k+1}^{n+1}.
\end{equation}
The step initial condition for TASEP particles is $z_1^n(0)=0$, $n\geq 1$. An analysis similar to the one of Section~\ref{subsectTASEPandGrowth} leads to $z_k^n(0)=k-1$, $1\leq k\leq n$. Then, the dynamics on $\{z_k^n,1\leq k\leq n, n\geq 1\}$ inherited by discrete time parallel update TASEP is the following. First of all, during the time-step from $n-1$ to $n$, all particles with upper-index greater or equal to $n+1$ are frozen. Then, from level $n$ down to level $1$, particles independently jump on their right site with probability $1/2$, provided the interlacing condition (\ref{eqInterAztec}) with the lower levels are satisfied. If the interlacing condition would be violated for particles in upper levels, then these particles are also pushed by one position to restore~(\ref{eqInterAztec}).

Finally, let us explain how to associate a tiling configuration to a particle configuration. For that we actually need to know the particle configuration at time $t=n$ and its previous time. Up to time $t=n$ only particles with upper-index at most $n$ could have possibly moved. These are also the only particles which are taken into account to determine the random tiling.
\begin{figure}
\begin{center}
\psfrag{z11}{$z_1^1$}
\psfrag{z12}{$z_1^2$}
\psfrag{z22}{$z_2^2$}
\psfrag{t0}[c]{$t=0$}
\psfrag{t1}[c]{$t=1$}
\psfrag{t2}[c]{$t=2$}
\includegraphics[width=\textwidth]{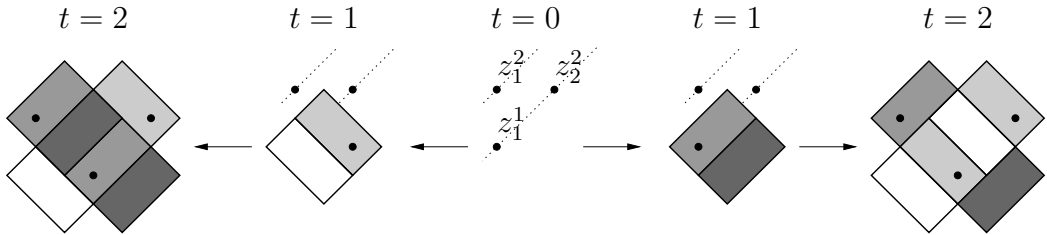}
\caption{Two examples of configurations at time $t=2$ obtained by the particle dynamics and its associated Aztec diamonds (rotated by 45 degrees).}
\label{FigAztec}
\end{center}
\end{figure}
The tiling follows these rules, see Figure~\ref{FigAztec} for an illustration:
\begin{enumerate}
  \item light-gray tiles: placed on each particle which moved in the last time-step,
  \item middle-gray tiles: placed on each particle which did not move in the last time-step,
  \item dark-gray tiles and white tiles: in the remaining position, depending on the tile orientation.
\end{enumerate}

The proof of the equivalence of the dynamics can be found in~\cite{Nor08}, where particle positions are slightly shifted with respect to Figure~\ref{FigAztec}. In~\cite{FerAZTEC} you can find a Java animation of the dynamics.

\appendix
\section{Further references}\label{AppAsymptRef}
In this section we give further references, in particular, of papers based on the approach described in this lecture notes.

\begin{itemize}
\item \emph{Interlacing structure and random matrices}: In~\cite{JN06} it is studied the GUE minor process which arises also in the Aztec diamond at the turning points. Turning points and GUE minor process occurs also for some class of Young diagrams~\cite{OR06}. The antisymmetric version of the GUE minors is studied in~\cite{FN08}. In~\cite{FN08b} the correlation functions for several random matrix ensembles are obtained, using two methods: the interlacing structure from~\cite{BFPS06} and the approach of~\cite{FN98}. When taking the limit into the bulk of the GUE minors one obtains the bead process, see~\cite{Bou09}. Further works on interlacing structures are~\cite{DW07,Jo07,MCW08,War07}.
\item \emph{GUE minors and TASEP}: Both the GUE minor process and the antisymmetric one occurs in the diffusion scaling limit of TASEP~\cite{BFS07,BFS09}.
\item \emph{$2+1$ dynamics}: The Markov process on interlacing structure introduced in~\cite{BF08} is not restricted to continuous time TASEP, but it is much more general. For example, it holds for PushASEP dynamics~\cite{BK07} and can be used for growth with a wall too~\cite{BK09}. In a discrete setting, a similar approach leads to a shuffling algorithm for boxed plane partitions~\cite{BG08}. As already mentioned, the connection between shuffling algorithm and interlacing particle dynamics is proven in~\cite{Nor08} (the connection with discrete time TASEP is however not mentioned).
\item \emph{$2+1$ anisotropic growth}: In the large time limit in the $2+1$ growth model the Gaussian Free Field arises, see~\cite{BF08} or for a more physical description of the result~\cite{BF08b}. In particular, height fluctuations live on a $\sqrt{\ln{t}}$ scale (in the bulk) and our model belongs to the anisotropic KPZ class, like the model studied in~\cite{PS97}.
\item \emph{Interlacing and asymptotics of TASEP}: Large time asymptotics of TASEP particles' positions with a few but important types of initial condition have been worked out using the framework initiated with~\cite{BFPS06}. Periodic initial conditions are studied in~\cite{BFPS06} and for discrete time TASEP (sequential update~\cite{BFP06}, parallel update~\cite{BFS07b}). The limit process of the rescaled particles' positions is the Airy$_1$ process. For step initial condition it was the Airy$_2$ process~\cite{Jo03b}. The transition process between these two has been discovered in~\cite{BFS07}, see also the review~\cite{Fer07}. Finally, the above technique can be used also for non-uniform jump rates where shock can occur~\cite{BFS09}.
\item \emph{Line ensembles method and corner growth models}: TASEP can be also interpreted as a growth model, if the occupation variables are taken to be the discrete gradient of an interface. TASEP belongs to the so-called Kardar-Parisi-Zhang (KPZ) universality class of growth models. It is in this context that the first connections between random matrices and stochastic growth models have been obtained~\cite{Jo00b}. The model studied is analogue to step initial conditions for TASEP. This initial condition can be studied using non-intersection line ensembles methods~\cite{Jo02b,Jo03b}. The Airy$_2$ process was discovered in~\cite{PS02} using this method, see also~\cite{Spo05,Jo05,Sas07} for reviews on this technique. The non-intersecting line description is used also to prove the occurrence of the Airy$_2$ process at the edge of the frozen region in the Aztec diamond~\cite{Jo03}.
\item \emph{Stationary TASEP and directed percolation}: Directed percolation for exponential/geometric random variables is tightly related with TASEP. In particular, the two-point function of stationary TASEP can be related with a directed percolation model~\cite{PS01}. The large time behavior of the two-point function conjectured in~\cite{PS01} based on universality was proven in~\cite{FS05a}. Some other universality-based of~\cite{PS01} conjectures have been verified in~\cite{BC09}. The large time limit process of particles' positions in stationary TASEP, the corresponding point-to-point directed percolation (with sources), and also for a related queuing system, has been unraveled in~\cite{BFP09}. The different models shares the same asymptotics due to the slow-decorrelation phenomena~\cite{Fer08}.
\item \emph{Directed percolation and random matrices}: Directed percolation, the Schur process and random matrices have also nice connections; from sample covariance matrices~\cite{BBP06}, to small rank perturbation of Hermitian random matrices~\cite{Pec05}, and to the generalization~\cite{BP07}.
\end{itemize}

\section{Christoffel-Darboux formula}\label{appChristoffel}
Here we prove Christoffel-Darboux formula (\ref{eqChristDarb}). First of all, we prove the three term relation (\ref{eq3TermRel}). From $q_n(x)/u_n=x^n+\cdots$ it follows that
\begin{equation}
\frac{q_n(x)}{u_n}-\frac{x q_{n-1}(x)}{u_{n-1}}
\end{equation}
is a polynomials of degree $n-1$. Thus,
\begin{equation}
\frac{q_n(x)}{u_n}=\frac{x q_{n-1}(x)}{u_{n-1}}+\sum_{k=0}^{n-1}\alpha_k q_k(x),\quad \alpha_k=\left\langle \frac{q_n}{u_n}-\frac{X q_{n-1}}{u_{n-1}},q_k\right\rangle_\omega,
\end{equation}
where $X$ is the multiplication operator by $x$, and $\langle f,g\rangle_\omega=\int_{\R} \omega(x) f(x) g(x)\dx x$ is the scalar product.

Let us show that $\alpha_k=0$ for $k=0,\ldots,n-3$. Using $\langle X f,g\rangle_\omega = \langle f,Xg\rangle_\omega$ we get
\begin{equation}
\alpha_k=\frac{1}{u_n}\langle q_n,q_k\rangle_\omega-\frac{1}{u_{n-1}}\langle q_{n-1},X q_k\rangle_\omega =0
\end{equation}
for $k+1<n-1$, since $X q_k$ is a polynomial of degree $k+1$ and can be written as linear combination of $q_0,\ldots,q_{k+1}$.

Consider next $k=n-2$. We have
\begin{equation}
\alpha_{n-2}=-\frac{1}{u_{n-1}}\langle q_{n-1},X q_{n-2}\rangle_\omega = -\frac{u_{n-2}}{u_{n-1}^2},
\end{equation}
because we can write
\begin{equation}
\begin{aligned}
x q_{n-2}(x) &= u_{n-2} x^{n-1} +\textrm{a polynomial of degree }n-2\\
&=\frac{u_{n-2}}{u_{n-1}} q_{n-1}(x) +\textrm{a polynomial of degree }n-2.
\end{aligned}
\end{equation}
Therefore, setting $B_n=\alpha_{n-1} u_n$, $A_n=u_{n}/u_{n-1}$, and $C_n=u_n u_{n-2}/u_{n-1}^2$, we obtain the three term relation (\ref{eq3TermRel}).
We rewrite it here for convenience,
\begin{equation}\label{eq3TermRelB}
q_n(x)=(A_n x+B_n)q_{n-1}(x)-C_n q_{n-2}(x).
\end{equation}
From (\ref{eq3TermRelB}) it follows
\begin{multline}\label{eq80}
q_{n+1}(x)q_n(y)-q_n(x)q_{n+1}(y)\\
=A_{n+1} q_n(x) q_n(y) (x-y)+ C_{n+1} \left(q_n(x) q_{n-1}(y)-q_{n-1}(x)q_n(y)\right).
\end{multline}
We now consider the case $x\neq y$. The case $x=y$ is obtained by taking the $y\to x$ limit.
Dividing (\ref{eq80}) by $(x-y)A_{n+1}$ we get, for $k\geq 1$,
\begin{equation}
q_k(x) q_k(y) = S_{k+1}(x,y)-S_{k}(x,y),
\end{equation}
where we defined
\begin{equation}
S_k(x,y)=\frac{u_{k-1}}{u_{k}}\frac{q_{k}(x)q_{k-1}(y)-q_{k-1}(x)q_{k}(y)}{x-y}.
\end{equation}
Therefore (for $x\neq y$)
\begin{equation}
\sum_{k=0}^{N-1} q_k(x)q_k(y) = S_{N}(x,y)-S_1(x,y)+q_0(x)q_0(y) = S_N(x,y).
\end{equation}
The last step uses $q_0(x)=u_0$ and $q_1(x)=u_1x+c$ (for some constant $c$), from which it follows $q_0(x) q_0(y)=S_1(x,y)$. This ends the derivation of the Christoffel-Darboux formula.

\section{Proof of Proposition~\ref{propGUEcorrfcf}}\label{AppGUEcorr}
Here we present the details of the proof of Proposition~\ref{propGUEcorrfcf} since it shows how the choice of the orthogonal polynomial is convenient. The basic ingredients of the proof of Theorem~\ref{thmBoro} are the same, with the only important difference that the functions in the determinants in (\ref{eqPropDPP}) are not yet biorthogonal.

First of all, let us verify the two relations (\ref{eq16}). We have
\begin{equation}
\int_{\R} K_N^{\rm GUE}(x,x)\dx x=\sum_{k=0}^{N-1} \langle q_k,q_k\rangle_\omega = N,
\end{equation}
and
\begin{equation}
\begin{aligned}
\int_{\R} K_N^{\rm GUE}(x,z)K_N^{\rm GUE}(z,y)\dx z&=\sum_{k,l=0}^{N-1} \sqrt{\omega(x) \omega(y)}q_k(x)q_l(y) \langle q_k,q_l\rangle_\omega\\
&=K_N^{\rm GUE}(x,y).
\end{aligned}
\end{equation}

By Lemma~\ref{lemGUEcorrfct}, Equation (\ref{eq13}), and the definition of $K_N^{\rm GUE}$, we have
\begin{multline}
\rho^{(n)}_{\rm GUE}(x_1,\ldots,x_n) \\
=c_N\frac{N!}{(N-n)!}\int_{\R^{N-n}}\det\left[K_N^{\rm GUE}(x_i,x_j)\right]_{1\leq i,j\leq N} \dx x_{n+1}\ldots \dx x_N.
\end{multline}
We need to integrate $N-n$ times, each step is similar. Assume therefore that we already reduced the size of the determinant to $m\times m$, i.e., integrated out $x_{m+1},\ldots,x_N$. Then, we need to compute
\begin{equation}\label{eq88}
\int_\R \det\left[K_N^{\rm GUE}(x_i,x_j)\right]_{1\leq i,j\leq m} \dx x_m.
\end{equation}
In what follows we write only $K$ instead of $K_N^{\rm GUE}$. We develop the determinant along the last column and get
\begin{equation}
\begin{aligned}
\det\left[K(x_i,x_j)\right]_{1\leq i,j\leq m}&= K(x_m,x_m) \det\left[K(x_i,x_j)\right]_{1\leq i,j\leq m-1}\\
&+\sum_{k=1}^{m-1} (-1)^{m-k} K(x_k,x_m)\det\left[
\begin{array}{c}
\left[ K(x_i,x_j)\right]_{\begin{subarray}{c}
                        1\leq i,j\leq m-1,\\ i\neq k
                          \end{subarray}}\\[1em]
\left[K(x_m,x_j)\right]_{1\leq j\leq m-1}
\end{array}
\right]\\
&= K(x_m,x_m) \det\left[K(x_i,x_j)\right]_{1\leq i,j\leq m-1}\\
&+\sum_{k=1}^{m-1} (-1)^{m-k} \det\left[
\begin{array}{c}
\left[ K(x_i,x_j)\right]_{\begin{subarray}{c}
                        1\leq i,j\leq m-1,\\ i\neq k
                          \end{subarray}}\\[1em]
\left[K(x_k,x_m) K(x_m,x_j)\right]_{1\leq j\leq m-1}
\end{array}
\right].
\end{aligned}
\end{equation}
Finally, by using the two relations (\ref{eq16}), Equation (\ref{eq88}) becomes
\begin{equation}
\begin{aligned}
&N \det\left[K(x_i,x_j)\right]_{1\leq i,j\leq m-1}
+\sum_{k=1}^{m-1} (-1)^{m-k} \det\left[
\begin{array}{c}
\left[ K(x_i,x_j)\right]_{\begin{subarray}{c}
                        1\leq i,j\leq m-1,\\ i\neq k
                          \end{subarray}}\\[1em]
\left[K(x_k,x_j)\right]_{1\leq j\leq m-1}
\end{array}
\right]\\
&=(N-(m-1))\det\left[K(x_i,x_j)\right]_{1\leq i,j\leq m-1}.
\end{aligned}
\end{equation}
This result, applied for $m=N,N-1,\ldots,n+1$, leads to
\begin{equation}
\rho^{(n)}_{\rm GUE}(x_1,\ldots,x_n)=c_N N! \det\left[K_N^{\rm GUE}(x_i,x_j)\right]_{1\leq i,j\leq n}.
\end{equation}
Now we need to determine $c_N$. Since $c_N$ depends only of $N$, we can compute it for the $n=1$ case. From the above computations, we have \mbox{$\rho^{(1)}_{\rm GUE}(x)=c_N N! K_N^{\rm GUE}(x,x)$} and $\int_\R \rho^{(1)}_{\rm GUE}(x)\dx x=N$ we have $c_N=1/N!$.

\end{document}